\documentclass[%
 reprint,
superscriptaddress,
nofootinbib,
 amsmath,amssymb,
 aps,
]{revtex4-2}

\usepackage[english]{babel}
\usepackage{tikz}
\usetikzlibrary{arrows.meta}
\usepackage{comment}
\usepackage{lipsum}
\usepackage{booktabs}
\usepackage{bbm}
\usepackage{graphicx}
\usepackage{dcolumn}
\usepackage{bm}
\usepackage{hyperref}

\begin{document}

\preprint{APS/123-QED}

\title{Thermodynamics of driven systems with explicitly broken detailed balance
}

\author{Markus Hofer}
\affiliation{Medical University of Vienna, Center for Medical Data Science, Institute of the Science of Complex Systems, Spitalgasse 23, 1090, Vienna, Austria}

 \affiliation{Complexity Science Hub Vienna, Josefst\"adterstrasse 39, 1080, Vienna Austria}
\author{Jan Korbel }
\affiliation{Medical University of Vienna, Center for Medical Data Science, Institute of the Science of Complex Systems, Spitalgasse 23, 1090, Vienna, Austria}

 \affiliation{Complexity Science Hub Vienna, Josefst\"adterstrasse 39, 1080, Vienna Austria}
\author{Rudolf Hanel}
\affiliation{Medical University of Vienna, Center for Medical Data Science, Institute of the Science of Complex Systems, Spitalgasse 23, 1090, Vienna, Austria}

 \affiliation{Complexity Science Hub Vienna, Josefst\"adterstrasse 39, 1080, Vienna Austria}
\author{Stefan Thurner}
\email{stefan.thurner@meduniwien.ac.at}
\affiliation{Medical University of Vienna, Center for Medical Data Science, Institute of the Science of Complex Systems, Spitalgasse 23, 1090, Vienna, Austria}

 \affiliation{Complexity Science Hub Vienna, Josefst\"adterstrasse 39, 1080, Vienna Austria}
 
\affiliation{ Santa Fe Institute, 1399 Hyde Park Rd, Santa Fe, NM 87501, USA
}%

\date{\today} 
             
\begin{abstract}
In systems with detailed balance, the stationary distribution and the equilibrium distribution are identical, creating a clear connection between energetic and entropic quantities. Many driven systems violate detailed balance and still pose a challenge for a consistent thermodynamic interpretation. Even steady-state potentials like entropy or free energy are no longer state variables. Here, we use a framework for systems with broken detailed balance, where Boltzmann entropy can be computed while properly taking constraints on state transitions into account. As an illustration, we establish the thermodynamic relations for arbitrarily driven sample space-reducing processes that are non-equilibrium but show steady states. We demonstrate that, despite explicitly broken detailed balance, it remains feasible to define and unambiguously interpret the effective thermodynamic potentials.
\end{abstract}

\maketitle
Physical systems with microscopic reversibility exhibit detailed balance, a condition of vanishing probability currents, linking energy and entropy through the second law of thermodynamics. 
This balance relates the transition rates of systems governed by a master equation,  
$p_i(t+1) = \sum_j T_{ij} p_j(t)$, to their equilibrium distribution.
Here,  $p_i(t)$  is the probability of being in state  $i$  at time  $t$, and  $T$  is the transition matrix. 
Detailed balance means  $T_{ij} \pi_j - T_{ji} \pi_i = 0$, i.e., forward and backward currents balance each other.
Consequently, after free relaxation, the stationary distribution  $\pi_i = \lim_{t \rightarrow \infty} p_i(t)$  becomes equilibrium distribution. 
This allows entropy change to be decomposed into heat flow and entropy production,  $\Delta S = \Delta Q / T + \Delta S_i$. 
Since entropy production is non-negative,  $\Delta S_i \geq 0$, we obtain the second law of thermodynamics,  $\Delta S \geq \Delta Q / T$ \cite{van2013stochastic}. 
Equality holds in equilibrium (or quasistatic) processes, where internal energy becomes a potential with entropy  $S$  as a state variable, making thermodynamic process properties path-independent.

However, most systems do not obey detailed balance. 
Consequently, thermodynamic potentials like internal or free energy become path-dependent (e.g., \cite{hanelTimeEnergyUncertainty2020}).
Dissipative systems driven between a source and a sink are examples of this behavior. 
These non-equilibrium conditions are essential for many biological \cite{fangNonequilibriumPhysicsBiology2019}, chemical \cite{polettiniIrreversibleThermodynamicsOpen2014}, and active matter systems \cite{gnesottoBrokenDetailedBalance2018}. Already, real gas molecules,
due to collision-induced dipole moments, do not collide fully elastically and maintain their kinetic energy only when in equilibrium with the thermal radiation they emit and absorb. 
Inelastic gases thus provide a simple physical example of this general situation
\cite{ben-naimStationaryStatesEnergy2005,thurnerEnergyDistributionInelastic2023}.
Systems that relax to a non-equilibrium steady state (NESS),
rather than to equilibrium, 
necessarily violate detailed balance. 
Non-equilibrium steady states can be analyzed through the concepts of housekeeping heat, 
needed to maintain the state, and excess heat, 
required for transitioning from one NESS to another \cite{oono1998steady,hatano2001steady}. 
Additionally, their entropy production can be divided into adiabatic and non-adiabatic parts \cite{PhysRevLett.104.090601}.
However, since entropy measured by the Shannon functional is no longer a state variable, thermodynamic potentials akin to those in equilibrium theory are not guaranteed.

In systems with detailed balance, maximizing Shannon entropy under constraints, 
e.g., a constraint on the internal energy yields the equilibrium distribution.
In a NESS, due to non-vanishing net probability flows, the stationary distribution no longer maximizes Shannon entropy under the same constraints. 
Thus, naively applying the maximum entropy principle \cite{jaynesInformationTheoryStatistical1957}
does not yield 
consistent
thermodynamic potentials. 
Therefore, a more general framework is needed to account for probability flows that break detailed balance.

Path-based approaches include the principle of Maximum Caliber \cite{jaynesMaxCal,RevModPhys.85.1115}, 
which measures trajectory entropy.
This principle is equivalent to Shannon’s information production, also known as entropy rate in coding theory \cite{10.1214/aoms/1177729028},
where it generalizes Shannon entropy. 
Despite this, it remains firmly grounded in Boltzmann entropy \cite{hanelHowMultiplicityDetermines2014} and serves as a starting point for constructing entropy functionals over the marginal state distribution function of the system \cite{hanelMaximumConfigurationPrinciple2018}, which can be achieved for a large class of processes \cite{hanel2011comprehensive}.

Such non-Shannonian functionals, often called generalized entropies, measure entropy based on the system’s state (the marginal distribution of states).
Examples of generalized entropies include Rényi entropy \cite{jizba2004world}, Tsallis entropy \cite{tsallis1988possible}, and $(c,d)$-entropies \cite{hanel2011comprehensive}. The motivation for these functionals is rooted in information-theoretic \cite{liese2006divergences,Renyi1976} and axiomatic reasons \cite{abe2000axioms}, or to capture fundamental scaling properties of stochastic systems \cite{hanel2011comprehensive}. They are sought for describing complex systems \cite{thurner2018introduction} or understanding the emergence of non-exponential distribution functions \cite{doi:10.1073/pnas.1320578110}.

Following Planck \cite{planck1901law}, we strictly adhere to Boltzmann’s prescription and define entropy as the logarithm of multiplicity, which is possible for a vast class of processes, at least asymptotically \cite{hanelTypicalSetEntropy2023}. 
Multiplicity is defined as the size of the typical set of trajectories (and thus entropy), while the ensuing asymptotic equipartition property, where trajectories in the typical set have essentially the same probability of occurring, identifies cross-entropy.
As we demonstrate below, thermodynamic relations, contrary to expectations from equilibrium theory, may naturally extend to dissipative systems steadily driven between a source and a sink.
For processes other than independent, identically distributed (IID) ones, multiplicity differs from the multinomial coefficient, resulting in non-Shannon types of entropy functionals \cite{thurner2017three}.

We focus on a process class known as sample-space reducing (SSR) processes \cite{thurnerSSR2015,corominas-murtraSampleSpaceReducing2017a}.
Besides being analytically tractable and being a statistical model of driven dissipative systems, they explicitly break the detailed balance, with driving and relaxation (dissipation) clearly disentangled. Dissipation progressively confines the system to smaller state space regions. By assigning monotonically decreasing energies to each state, the system relaxes towards its thermal ground state. 
Conversely, with a given rate, the system is randomly reset to states with possibly higher energy, expanding the state space volume. Without driving, the system cannot be ergodic, as states cannot be revisited under
dissipation alone. The interplay between driving and dissipation results in a NESS, where the steady-state distribution follows an exact power law for a fixed driving rate \cite{corominas-murtraSampleSpaceReducing2017a}.

SSR processes have been successfully used to explain why word frequencies in natural language follow Zipf’s law \cite{thurnerUnderstandingZipfLaw2015, mazzoliniHeapsLawStatistics2018}, 
to understand visiting frequencies in diffusion processes on directed acyclic graphs, statistics of fragmentation processes \cite{corominas-murtraSampleSpaceReducing2017a}, and the energy (and velocity) distribution of inelastic gases \cite{thurnerEnergyDistributionInelastic2023}. The entropy and cross-entropy of generally driven SSR processes were derived in \cite{hanelMaximumConfigurationPrinciple2018}.
Here, we concisely interpret the ensuing terms as the respective thermodynamic equations that extend the thermodynamic potentials from quasi-static equilibrium theory to NESS, arbitrarily far from equilibrium.

As is the case for equilibrium thermodynamics, the Legendre transformations still connect internal energy, entropy, inverse temperature, and free energy of NESS systems that violate detailed balance.
All the thermodynamic quantities (entropy, internal energy, and the (non-equilibrium version of) Helmholtz free energy)  remain thermodynamic potentials, where, e.g., $S$ is again a macro-state variable.

With vanishing driving rates, the correction terms, which distinguish NESS quantities from their equilibrium versions, also vanish. Consequently, these correction terms represent generalized forces that determine how far the process operates from equilibrium and how strongly the system violates detailed balance.

\emph{Entropy via Boltzmann's principle. ---} Boltzmann's principle identifies the entropy of a system as the log-multiplicity of its underlying sample space volume and the cross entropy as the logarithm of the probability of a typical sequence of states. It can be extended to general stochastic processes as $ H = \lim_{n \rightarrow \infty} \frac{1}{n} \log P(x_n,x_{n-1},\dots,x_1) $
where $P(x)$ is the probability of observing a trajectory, 
$x=(x_1,x_2,\cdots,x_{n-1},x_n)$ \cite{10.1214/aoms/1177729028} . 

We exemplify Boltzmann’s principle with a simple IID process.
Suppose our initial knowledge about a system is limited to its number of possible states,  $W$.
To assign a probability to each state, we consider how often a state is observed during a trajectory with $N \gg W$ 
consecutively measured states. If the number of times state  $i$  occurred is  $k_i$ , the probability of observing a specific distribution of state visits  $k = (k_1, k_2, \ldots, k_W)$  is given by
\begin{equation}
        P(k \, \vert \,  q) = 
    \underbrace{
        \frac{N !}{k_{1} ! k_{2} ! \ldots k_{\lvert \Omega \rvert} !}
    }_{M(k)}
    \;
    \underbrace{
        q_1^{k_1} q_2^{k_2} \ldots 
        q_{\lvert \Omega \rvert}^{k_{\lvert \Omega \rvert}}
    }_{G(k \, \vert \,  q)} \, ,
    \label{eqn:MultinomialMult}
\end{equation}
where $q_i$ represents the probability of sampling state $i$ \cite{hanelHowMultiplicityDetermines2014}. 
Eq. \eqref{eqn:MultinomialMult} is a product of a multinomial $M(k)$ that counts the number of configurations with a specific distribution, $k$, 
and the probability of that configuration given by $G(k \,\vert \, q)$.

Using $p_i = k_i /N$ and Stirling's formula for large $N$, the normalized logarithm of multiplicity, $\log M(k)/N$ results in Shannon entropy $H(p) = - \sum_{i =1} ^{W} p_i \log p_i$. Similarly, we define the cross entropy as $ -\log G(k  \, \vert \,  q)/N$, which yields $H(p\, \vert \, q) = - \sum_{i= 1} ^{W}p_i \log q_i $. Finding the most likely distribution, $p_i$, i.e., the normalized histogram of state visits, is therefore equivalent to 
minimizing the Kullback-Leibler divergence 
$D_{\textrm{KL}}(p \, \vert\vert \, q) = H(p \, \vert \, q)  - H(p)$. 
By parameterizing the prior distribution, for example, as $ q_i = \exp{(-\alpha - \beta \epsilon_i)} $, the connection to equilibrium thermodynamic potentials is provided by the Helmholtz free energy
\begin{eqnarray}
    \label{eqn:equilibrium_potentials}
  \beta^{-1} D_{\textrm{KL}}(p \, \vert\vert \, q) &=& \langle \epsilon \rangle_{p} - \beta^{-1} H(p) + \beta^{-1} \alpha \nonumber\\
  &=& F(p) - F(q)\, , 
\end{eqnarray}
with $F(q) = - \beta^{-1} \alpha$ is the equilibrium free energy.
Where we use the notation  $\langle x \rangle_p \equiv \sum_{i=1}^W x_i p_i$  to denote the expected value of  $x$  with respect to the probability distribution  $p$. 

\begin{figure}
    \centering
    \includegraphics[width=0.6\columnwidth]{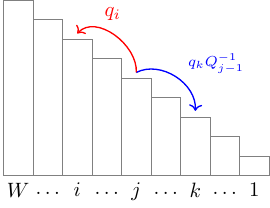}
    \caption{Illustration of transitions in a sample space reducing process with arbitrary driving. The blue arrow indicates an SSR step that can only move to lower states, $k<j$ (relaxation or dissipative step). The red arrow represents a driving step that can move the system to any state.}
    \label{fig:ssr-staircase}
\end{figure}

\emph{Sample-space reducing processes. ---}
We consider a general form of Markovian sample space-reducing processes \cite{thurner2017three}, which sample
a discrete, ordered state space $\Omega$ of size $W$. If the system is currently in state $j$, its next state is determined by two events. 
Either, with probability $1-\lambda_j \in [0,1]$, the process undergoes a driving event. In Fig. \ref{fig:ssr-staircase}, this is represented by a red arrow. In this case the next state, $i$, is sampled with prior probabilities, $q_i$.
This driving update can model the interaction of a system with a source, e.g., a particle reservoir or heat bath, in which case the prior probabilities, $q_i$, are set to the Boltzmann distribution. Alternatively, with probability $\lambda_j$, the process takes an SSR step, illustrated by a blue arrow in Fig \ref{fig:ssr-staircase}. 

Assuming that the states  $i = 1, 2, \cdots, W$  are ordered, typically in terms of their respective energies, the system moves to a new state  $k$  that is lower than  $j$  with the renormalized probabilities $q_k Q_{j-1}^{-1}$, where $Q_{j-1} \equiv \sum_{s=1}^{j-1} q_s$. This 
describes
the system’s relaxation, meaning energy (or particles) dissipates into a sink. 

Such a driven SSR process can be modeled using a discrete Markov chain with the transition matrix $T_{ij} = q_i +  \lambda_j q_i\left(\frac{  \Theta(j-i) }{Q_{j-1}}  -1 \right)$,
where $\lambda_j$  and  $q_i$  denote the same quantities discussed above. The  $q_i$ need not be normalized, but for simplicity, we assume  $\sum_i q_i = 1$. The Heaviside function,  $\Theta$, ensures that only transitions to lower states are allowed during an SSR event. 
From the transition matrix, we can determine the stationary distribution directly as
  $\pi_i = \frac{1}{\zeta} q_i \prod_{j=2}^{i} \left(1 + \lambda_j \frac{q_j}{Q_{j-1}} \right)^{-1}$, with the normalization constant $\zeta$. Since SSR steps progressively reduce the number of accessible states, the process could get stuck in the ground state,  $i = 1$. To avoid this scenario, we always set  $\lambda_1 = 0$. All other driving rates can be chosen freely.

The transition structure of SSR processes provides us with a well-defined multiplicity, $M$, for all trajectories $x$ that share the same histogram $k$ and the probability $G$ of any particular path with histogram $k$ (which, for the SSR case, are all identical). 
This multiplicity factor, however, is no longer a multinomial but
\begin{equation}
    \label{eqn:s_mc_paper}
   M(k, k^*) \equiv  \binom{\sum_{s=1}^{W}k_s^* }{k^*} \prod_{j=2}^{ W} 
    \binom{\sum_{s=1}^{j-1}k_s^*}{k_j - k_j^*} \, . 
\end{equation}

The detailed derivation can be found in the SI. $k$ denotes the distribution of states visited by the entire process and $k^*$ is the distribution of those visits immediately following a driving event.

Similarly, the probability, $G$, of observing a path that has exactly histogram $k$, gives 
\begin{equation}
    \label{eqn:s_cross_paper}
    G_\lambda(k, k^* \mid q) \equiv q_1^{k_1^*}  \prod_{i=2}^{ W }\left(\frac{\lambda_i q_i}{Q_{i-1}}\right)^{k_i-k_i^*}
     \prod_{j=2}^{ W }\left(\left(1-\lambda_j\right) q_j\right)^{k_j^*} \, .
\end{equation}
The derivation can be found in the SI. 

\emph{Entropy and cross-entropy for SSR processes. ---}
To derive the entropy functionals from Eq. \eqref{eqn:s_mc_paper}, we follow the same approach as we did for IID processes. 
First we define $ H(p, p^*) =  \lim_{N \rightarrow \infty} N^{-1} \log M(k,k^*) $, where $p_i = k_i/N$ and  
$p_i^* = k_i^*/N = (1- \lambda_i) p_i$ is the probability that a driving event occurs while the process is in state $i$. This allows us to re-express the entropy as a function of the driving rate and $p$
\begin{equation}
    H_{\lambda}(p) = H(p) + H\left(C -  p \right)  - H\left(C \right)  + \langle H_2(\lambda)\rangle_p \, , 
    \label{eqn:simple_gen_entropy}
\end{equation}
where $H(p)$ refers to Shannon entropy and $C_i=\sum_{k=1}^{i} (1-\lambda_k) p_k$
is the probability that a driving event occurs in any state below or equal to $i$. 
\footnote{
Note that in \cite{hanelMaximumConfigurationPrinciple2018}, a different normalization with respect to the number of driving events was employed.}
See SI for details.
The third term is the average {\it expected driving entropy}, 
$\langle H_2(\lambda)\rangle_p = - \sum_i p_i \left( \lambda_i \log \lambda_i + (1-\lambda_i) \log (1-\lambda_i) \right)$, 
associated with 
the choices made between driving events and sample space-reducing events in the process.

We calculate the cross entropy of an SSR process equivalently to the cross entropy for an IID process by defining
$H_\lambda(p, \; p^* \mid q) \equiv - \lim_{N\rightarrow \infty}  N^{-1} \log G_\lambda(k,k^* \mid q)$.
After taking the limit and inserting $p_i$ and $p_i^*$ as well as the relation between them, we obtain
\begin{equation}
    \label{eqn:simple_gen_cross_entropy}
    H_{\lambda}^{\textrm{cross}}(p \mid q) = H(p \mid q) -  \sum_{i=2}^W p_i \lambda_i \log (Q_{i-1})
    + \langle H_2(\lambda)\rangle_p \, .
\end{equation}
See SI for details. 

For both entropy and cross-entropy, there is a clear separation into two parts: a path-independent part with detailed balance (Shannon entropy and cross-entropy) and additional terms that account for the broken detailed balance. 
Note that in the limit of fast driving ($\lambda_i \to 0$), the terms incorporating the transition structure vanish, and the usual expressions for entropy and cross-entropy are recovered. This is expected since, in this limit, SSR processes regain detailed balance and, as such, follow equilibrium thermodynamics. See SI for details. 

Combining entropy end cross-entropy, we obtain a definition of divergence
\begin{equation}
    D_{\lambda} (p \, \vert \vert \,    q) \equiv H_{\lambda}^{\textrm{cross}}(p \; \vert q) - H_{\lambda}(p)  \, , 
    \label{eqn::div}
\end{equation}

Note that this divergence is minimized by the distribution function $p=\pi$, where $\pi\neq q$. See SI or \cite{hanelMaximumConfigurationPrinciple2018} for further details.
However, since $q$ uniquely determines a minimizer, $\pi$, we could define $\hat D_\lambda(p||\pi) \equiv D_\lambda(p||q(\pi))$.
The functional $\hat D_\lambda$ then has the familiar property $\hat D_\lambda( p||\pi)=0$ for $p=\pi$.

\emph{Thermodynamics of SSR processes. ---}
To understand the thermodynamics of systems following SSR processes, we parameterize the prior weights according to the Boltzmann distribution, $q_i = \exp(- \alpha - \beta \epsilon_i)$, with normalization $\alpha$. 
This means we assume our system to be in contact with an external heat bath at inverse temperature $\beta$ as a source of energy (or particles), while the dynamics within the system is restricted by the way the system dissipates energy (or particles) into a sink, which manifest in the form of SSR transitions.

This approach, when used for Shannon entropy and cross-entropy with this exact parameterization, yields the classic equilibrium thermodynamic potentials; see Eq. \eqref{eqn:equilibrium_potentials}.
However, due to its non-equilibrium nature, for the SSR process, we obtain additional terms,
\begin{eqnarray}
    \label{eqn:nonequilibrium_potentials}
  \beta^{-1} D_\lambda (p \, \vert\vert \, q) &=&  F(p) - F(q) \\ \nonumber
     - \langle \lambda(F&-&f) \rangle_p + \beta^{-1}\left( H(C) - H(C-p) \right) \, ,
\end{eqnarray}
where $f_i$ is the equilibrium free energy of a system composed of all states lower than $i$, i.e., $f_i = - \beta^{-1} \log \sum_{j=1}^{i-1} e^{-\beta \epsilon_j}$ and $F = - \beta^{-1} \log \sum_{j=1}^{W} e^{-\beta \epsilon_j}$ is the Helmholtz free energy of the unrestricted state space. Additionally we use $\langle  \lambda x \rangle_p \equiv \sum_{i=1}^W \lambda_i x_i p_i$. 

In the stationary state, we obtain the following relation between the NESS
thermodynamic potentials
\begin{equation}
    \label{eqn:effective_potentials}
    F_\lambda = U_\lambda  - \beta^{-1} S_\lambda \, .
\end{equation}

Here, $S_\lambda \equiv H_\lambda(\pi) - \langle H_2 ( \vec{\lambda}) \rangle_\pi$ represents the entropy of the state trajectories alone, as the average entropy associated with the choice between a driving step and a sample space-reducing step has been subtracted. These trajectories therefore follow a hidden Markov process, which means the process itself is no longer Markovian. As a result, $S_\lambda$ is lower than the conditional entropy of the trajectories.
The internal energy for the NESS state $\pi$ is simply given by $U_\lambda \equiv \langle \epsilon\rangle_\pi$. 

To see the smooth transition between NESS and equilibrium states with $\lambda\to 0$, the respective
non-equilibrium free energy $F_\lambda$ can be 
decomposed: 
\begin{equation}
    F_\lambda = F - \langle \lambda (F- f) \,\rangle_\pi \, .
\end{equation}

This also means that the non-equilibrium free energy in Eq. \eqref{eqn:effective_potentials} can be fully understood in terms of 
hierarchy of nested state spaces $\Omega_i  \equiv \{1,2,\cdots,i\}$ and their respective 
equilibrium free energies weighted by $\lambda$.

Similarly, we can decompose
$U_\lambda$ and 
$S_\lambda$:
\begin{equation}
   U_\lambda =  U -\langle \lambda (U-u) \rangle_\pi \qquad S_\lambda =  S -\langle \lambda (S-s)\,\rangle_\pi 
\end{equation}
where $S=H(q)$ with $q=\exp(-\alpha-\beta\epsilon_i))$ and $U=\langle \epsilon_i\rangle_q$.

The vector $u$ consists of the respective hierarchy of partial internal energies, $u_i = \sum_{j=1}^i \epsilon_j q_j/Q_{i-1}$. 
Similarly, $s$ contains the partial equilibrium entropies 
$s_i = -\sum_{j=1}^{i-1}  (q_j/Q_{i-1}) \log (q_j/Q_{i-1})$; see SI for the derivation.

The fact that $S_\lambda$ follows an extremal principle and
the Legendre transformation, Eq. \eqref{eqn:effective_potentials}, connects 
$S_\lambda$, $F_\lambda$, $U_\lambda$, and $\beta$
for all possible choices of $\lambda$, implies that $U_\lambda$ is a potential with state variable $S_\lambda$, which the Legendre transform takes into the potential $F_\lambda$ with state variable $1/\beta$.

We note, as an example, elastic collisions between a moving particle (projectile) and a stationary particle (target) of the same size and weight in three dimensions follow an SSR step with uniformly distributed weights $q$ \cite{thurnerEnergyDistributionInelastic2023}. In a system where a particle source emits projectiles with an energy distribution q, which scatter at several layers of resting targets (i.e., targets kept at a low temperature by a cold heat bath). A fast particle will scatter at each target layer, performing one SSR step per collision. Here, the number of target layers corresponds to the parameters $\lambda_i$, controlling the average number of SSR steps before the next driving event (next projectile emission). The layered target structure resembles hierarchies of partial internal energies u, entropies s, and free energies f.

Inelastic gases can be understood similarly \cite{thurnerEnergyDistributionInelastic2023}. Fast particles drive inelastic collisions, dissipating energy as heat into a thermal sink, with occasional energy added back. The ratio of collision frequency to energy addition translates into $\lambda$. However, the nested SSR hierarchy is complex, as slow particles act as targets and fast ones as projectiles, making the distribution $q$ dependent on particle velocity distribution.

\emph{Discussion. ---} We propose bridging the gap between traditional thermodynamics and systems far from equilibrium, where detailed balance does not hold. By adapting Boltzmann’s formula for state transition constraints in non-equilibrium processes, we established a framework using driven sample-space reducing processes to understand the thermodynamics of these systems.

SSR processes can describe systems operating between sources and sinks. These include abstract processes like language (grammatical and content constraints) \cite{thurnerUnderstandingZipfLaw2015}, search and directed diffusion on networks \cite{corominas-murtraExtremeRobustnessScaling2016}, fragmentation processes \cite{corominas-murtraSampleSpaceReducing2017a}, and record statistics \cite{yadavSurvivaltimeStatisticsSample2016a}. Physical collision processes also exhibit SSR characteristics, giving physical meaning to entropy, free energy, and inverse temperature, which connects them 
continuously
to their equilibrium thermodynamic counterparts.

Our work provides a more comprehensive understanding of NESS and offers new insights into the relationship between energetic and entropic quantities in physical systems influenced by driving and dissipative forces.
We demonstrate that non-equilibrium quantities define state variables in a non-equilibrium state, maintaining their roles as thermodynamic potentials connected through Legendre transformations, just as in equilibrium systems.

Notably, the entropy formulation presented here meets all three hypothesized criteria identified by physical entropy \cite{jaynesInformationTheoryStatistical1957}: it is mathematically sound, maintaining a clear connection to Boltzmann’s principle and the probability of observing distribution  $p$  in the system. It avoids arbitrary assumptions, incorporating only information about the system without requiring additional parameters. Most significantly, it includes both non-equilibrium and equilibrium systems.

One possible explanation for this approach’s effectiveness is as follows 
(see \cite{hanelEquivalenceInformationProduction2023} for further discussion): By adapting multiplicity and assignment probability to the sample-space reducing process, we implicitly construct an IID process within a larger state space. The elements of this extended state space consist of all decreasing sequences on the original state space. Since the resulting process in the extended state space is IID, its entropy can be accurately described by the Shannon entropy formula. However, to express this entropy in terms of the probabilities on the original state space, additional terms are needed in the entropy calculation.

More importantly, re-expressing path-based entropy concepts, like information production, in terms of functionals over the marginal distribution function allows for a smooth generalization of standard thermodynamic quantities while maintaining their role as thermodynamic potentials. Whenever a process can be expressed as an IID process in a larger state space, it necessitates using generalized entropy functionals to represent entropy production in the original state space accurately. Consequently, this also requires generalized potentials for the correct thermodynamic analysis of the system in a non-equilibrium steady state.

\begin{acknowledgments}
    We acknowledge support from the Austrian Science Fund, grant-doi:10.55776/P34994
\end{acknowledgments}

\end{document}